\documentclass[]{cupconf}
\usepackage{amssymb,graphicx,natbib}

\newcommand\teff{T_{\rm eff}}
\newcommand\kms{\rm\,km\,s^{-1}}

\newcommand\ulsr{U_{\rm LSR}}
\newcommand\vlsr{V_{\rm LSR}}

\title[Hercules stream stars and the metal-rich thick disk]
      {Hercules stream stars and the \\ metal-rich thick disk}
      
\author[T.~Bensby et al.]{\\ T.~Bensby$^1$, M.S.~Oey$^1$, S.~Feltzing$^2$
\& B.~Gustafsson$^{3}$}
\affiliation{$^1$ Department of Astronomy, University of Michigan, 
830 Dennison Bldg, 500 Church Street, Ann Arbor, MI~48109-1042, USA 
({\tt tbensby@umich.edu, msoey@umich.edu})\\ $^2$ Lund Observatory, Box 43, 
SE-22100 Lund, Sweden ({\tt sofia@astro.lu.se}) \\ $^3$ Department of 
Astronomy and Space Physics, University  of Uppsala, Box 515, \\
SE-75120 Uppsala, Sweden ({\tt bengt.gustafsson@astro.uu.se}) }   

\begin{document}
\maketitle

\begin{abstract}
Using the MIKE spectrograph, mounted on the
6.5\,m Magellan/Clay telescope at the Las Campanas observatory in
Chile, we have obtained high-resolution spectra for 60 F and G dwarf
stars, all likely members of a density enhancement in the local
velocity distribution, referred to as the Hercules stream.
Comparing with an existing sample of 102 thin and thick disk
stars we have used space velocities, detailed elemental abundances, and 
stellar ages to trace the origin of the Hercules stream.
We find that the Hercules stream stars show a wide spread in stellar ages,
metallicities, and element abundances. However, the spreads are not 
random but separate the Hercules stream into the abundance and age trends 
as outlined by either the thin disk or the thick disk. We hence claim that 
the major part of the Hercules stream actually are thin and thick disk stars.
These diverse properties of the Hercules stream point toward a dynamical 
origin, probably caused by the Galactic bar. However, we can at the
moment not entirely rule out that the Hercules stream could be the
remnants of a relatively recent merger event.
\end{abstract}

\firstsection
\section{Introduction}
The stellar velocity distribution in the solar neighbourhood
is not smooth, but shows lots of substructure
\citep[e.g.][]{dehnen1998a,skuljan1999,famaey2005,helmi2006,arifyanto2006}.
The most prominent features are the Pleiades-Hyades super-cluster,
the Sirius cluster, and the Hercules stream (also known as the
$u$-anomaly). Studying nearby G and K giants \cite{famaey2005}
found that the Hercules stream makes up $\sim 6$\,\% of the stars in the
solar neighbourhood with stars moving on highly eccentric orbits. 
On average these
stars have a net drift of $\sim 40\,\kms$ directed radially away from the
Galactic center, and just as for the thick disk, their orbital velocities
around the Galaxy lag behind the local standard of rest (LSR) by
$\sim 50\,\kms$ (see also \citealt{ecuvillon2006astroph} who found 
similar properties from nearby F and G dwarf stars).

As several numerical simulations have shown, this excess of stars at
$(\ulsr,\vlsr)\approx(-40,\,-50)\,\kms$ can be
explained as a signature of the Galactic bar
\citep[e.g.,][]{raboud1998,dehnen1999,dehnen2000,fux2001}.
Whether or not it is a chaotic process, where stars get gravitationally
scattered off the inner regions by the bar, or whether they have ordered orbits
coupled to the outer Lindblad resonance of the bar is, however, uncertain
\citep{fux2001}. In any case,
this explanation means that the stars in the Hercules stream should originate
from the inner disk regions. 
So, is the Hercules stream a distinct Galactic stellar population 
with a unique origin and history or is it a mixture of the other populations?
Or could they even be a remnant of an ancient merger event?

To further trace the origin of the Hercules stream
we have observed a sample
of 60 F and G dwarf stars. By performing a strictly differential
detailed abundance analysis of the Hercules stream stars relative to
stars of the two disk populations previously studied by us 
\citep{bensby2003,bensby2005}
we minimise uncertainties due to systematic errors in the analysis.

\begin{figure}
\centering
\resizebox{\hsize}{!}{\includegraphics[bb=18 154 592 520,clip]{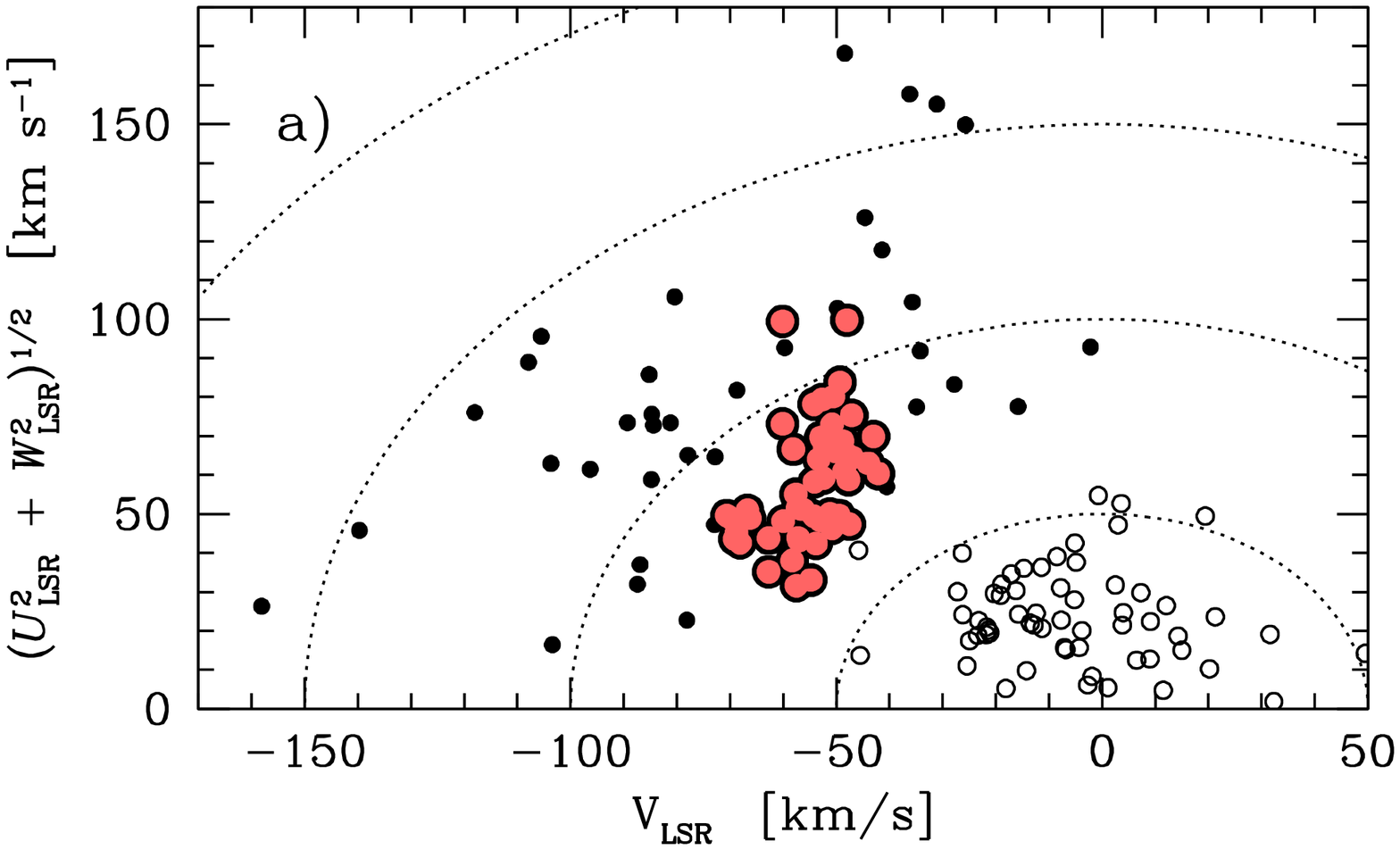}
	\includegraphics[bb=18 154 592 520,clip]{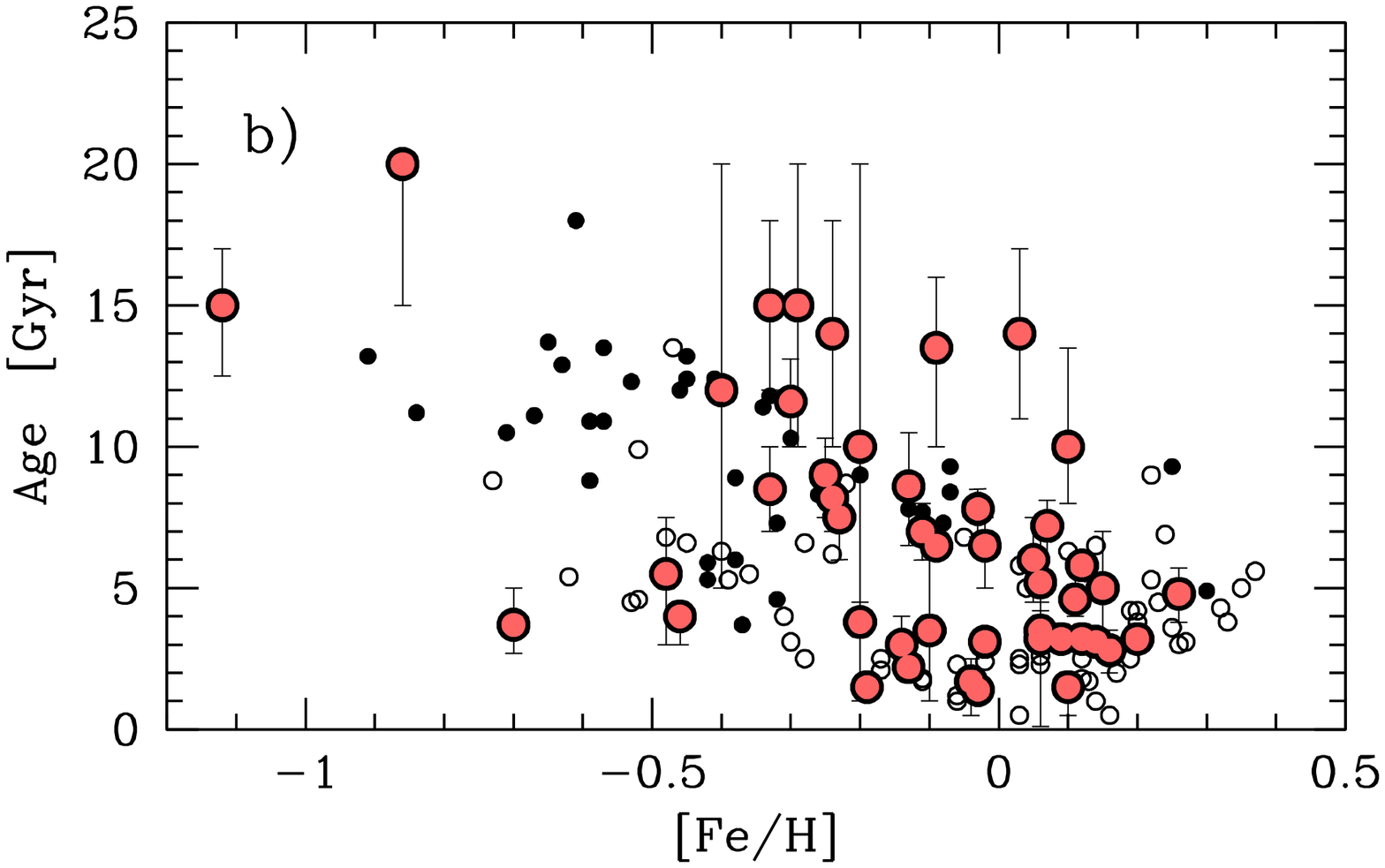}}
\caption{
	{\bf a)} Toomre diagram. 
	{\bf b)} Stellar ages versus [Fe/H].
	Hercules stream stars are marked by
	red (gray in B/W) stars, 
	thin and thick disk stars from \cite{bensby2003,bensby2005} 
	by open and black filled circles, respectively. Error bars are 
	explained in the text. 
        }
\label{fig:toomre}
\end{figure}

\section{Observations, abundance analysis, and age determinations}
\label{sec:observations}

High-resolution ($R\approx65\,000$), high-quality ($S/N\gtrsim250$)
echelle spectra were obtained for 60 F and G dwarfs by TB in Jan, Apr,
and Aug in 2006 with the MIKE
spectrograph \citep{bernstein2003} on the Magellan Clay 6.5\,m telescope at
the Las Campanas Observatory in Chile. A Toomre diagram for the sample
can be seen in  Fig.~\ref{fig:toomre}a.

For the abundance analysis we have used the Uppsala MARCS stellar model
atmospheres \citep{gustafsson1975,edvardsson1993,asplund1997}. These models
are one-dimensional, plane-parallel, and calculated under the assumption of
local thermodynamic equilibrium (LTE). Their chemical compositions have been
scaled with metallicity relative to the standard solar abundances as given in
\cite{asplundgrevessesauval2005}, using enhanced abundances for the
$\alpha$-elements at sub-solar metallicities. To determine effective
temperatures we use excitation equilibrium of Fe\,{\sc i}
and to estimate the microturbulence parameter we require all
Fe\,{\sc i} lines to yield the same abundance independent of line strength.
To estimate the surface gravities we utilise that all our stars have accurate
Hipparcos parallaxes \citep{esa1997}. 
Final abundances were first normalised on a line-by-line
basis with our solar values as reference and then averaged for each element.

Stellar ages were determined from the Yonsei-Yale (Y$^{2}$) $\alpha$-enhanced
isochrones \citep{kim2002,demarque2004} in the $\teff$-$M_{\rm V}$ plane.
Lower and upper limits on the ages were estimated by the error bars arising
from an assumed uncertainty of $\pm 70$\,K \citep[see][]{bensby2003} in
$\teff$ and the uncertainty in $M_{\rm V}$ due to the
error in the parallax.

\begin{figure*}
\resizebox{\hsize}{!}{\includegraphics[bb=18 223 562 515,clip]{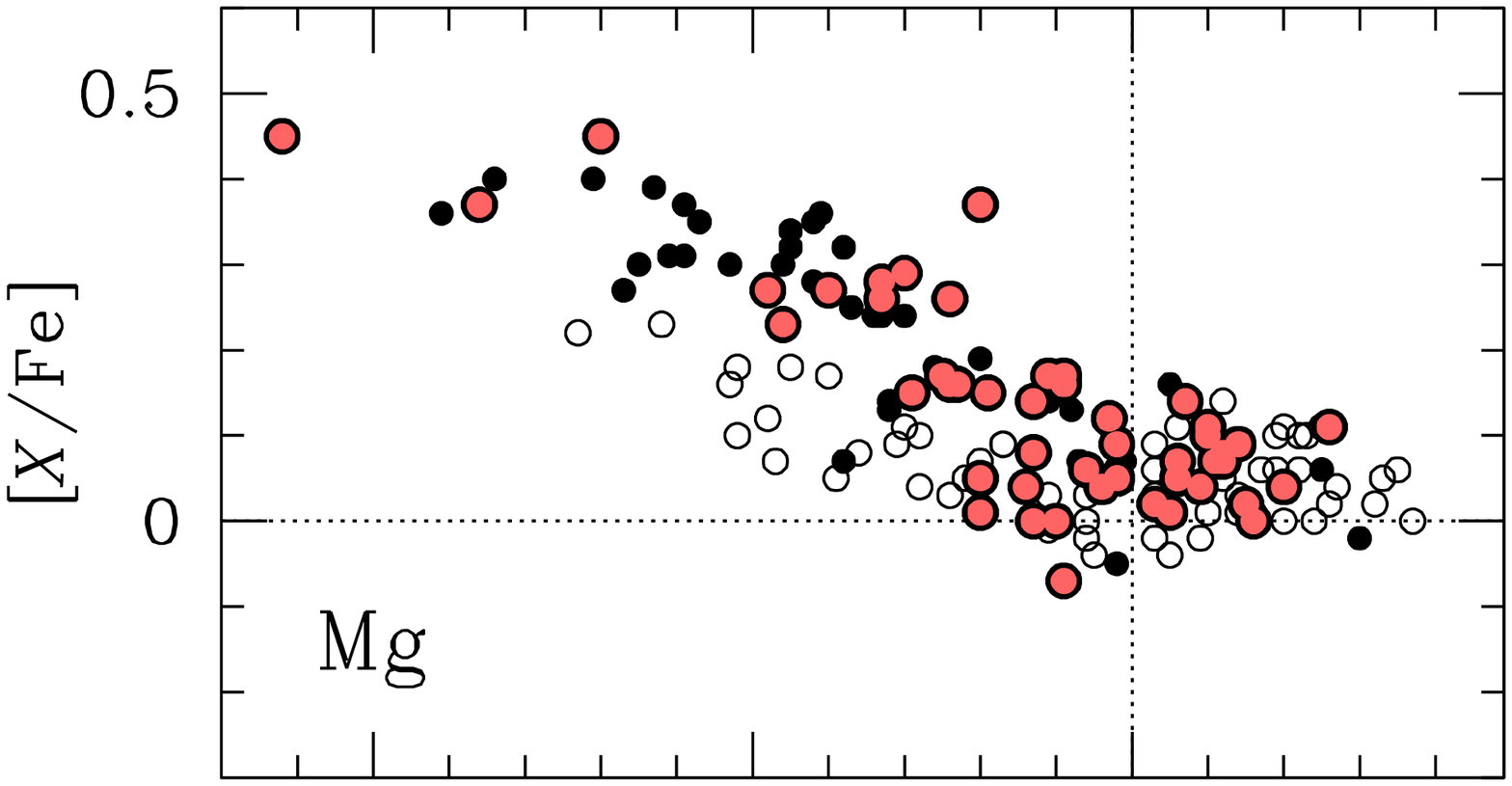}
                      \includegraphics[bb=90 223 562 515,clip]{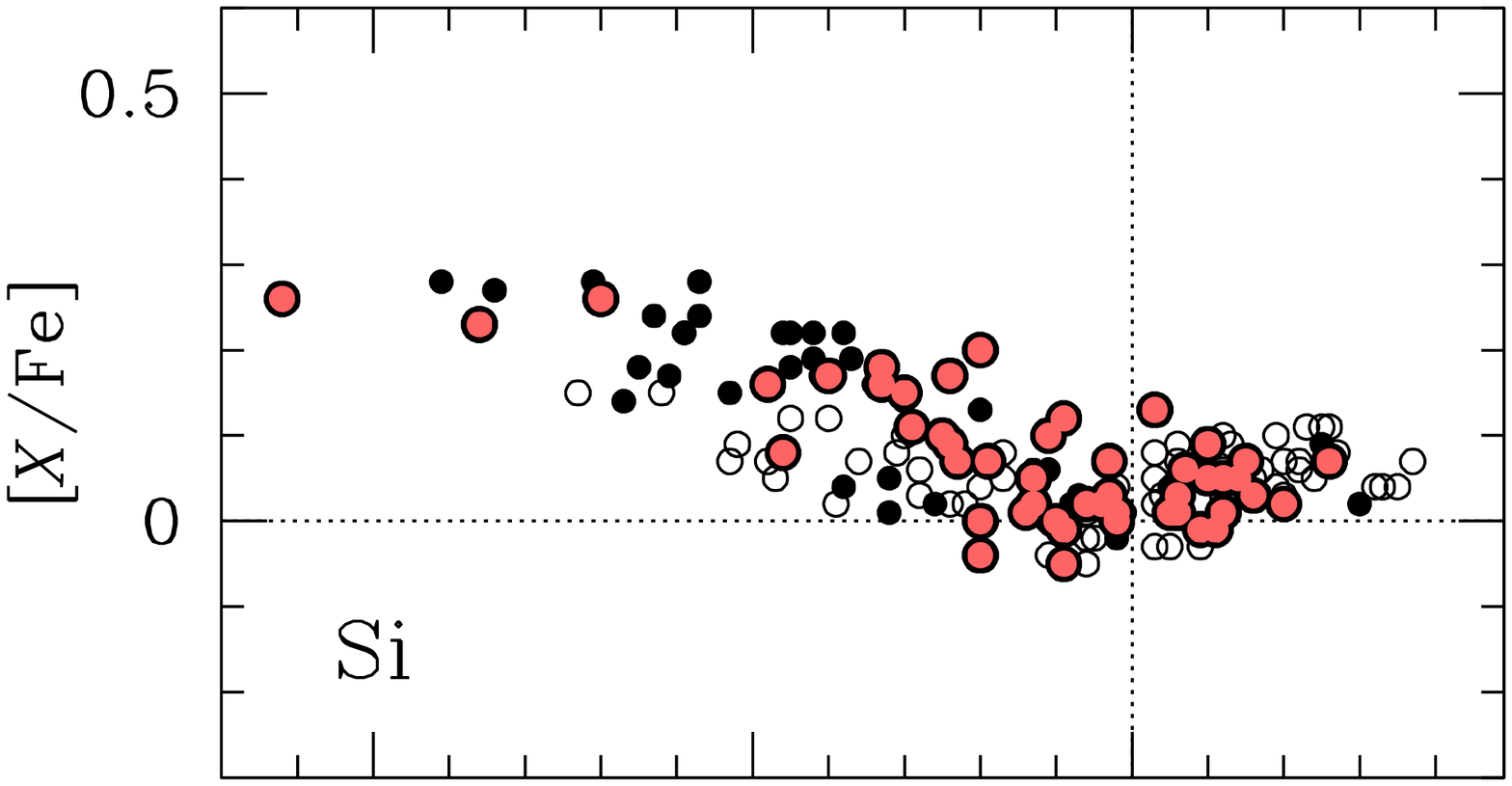}
                      \includegraphics[bb=90 223 592 515,clip]{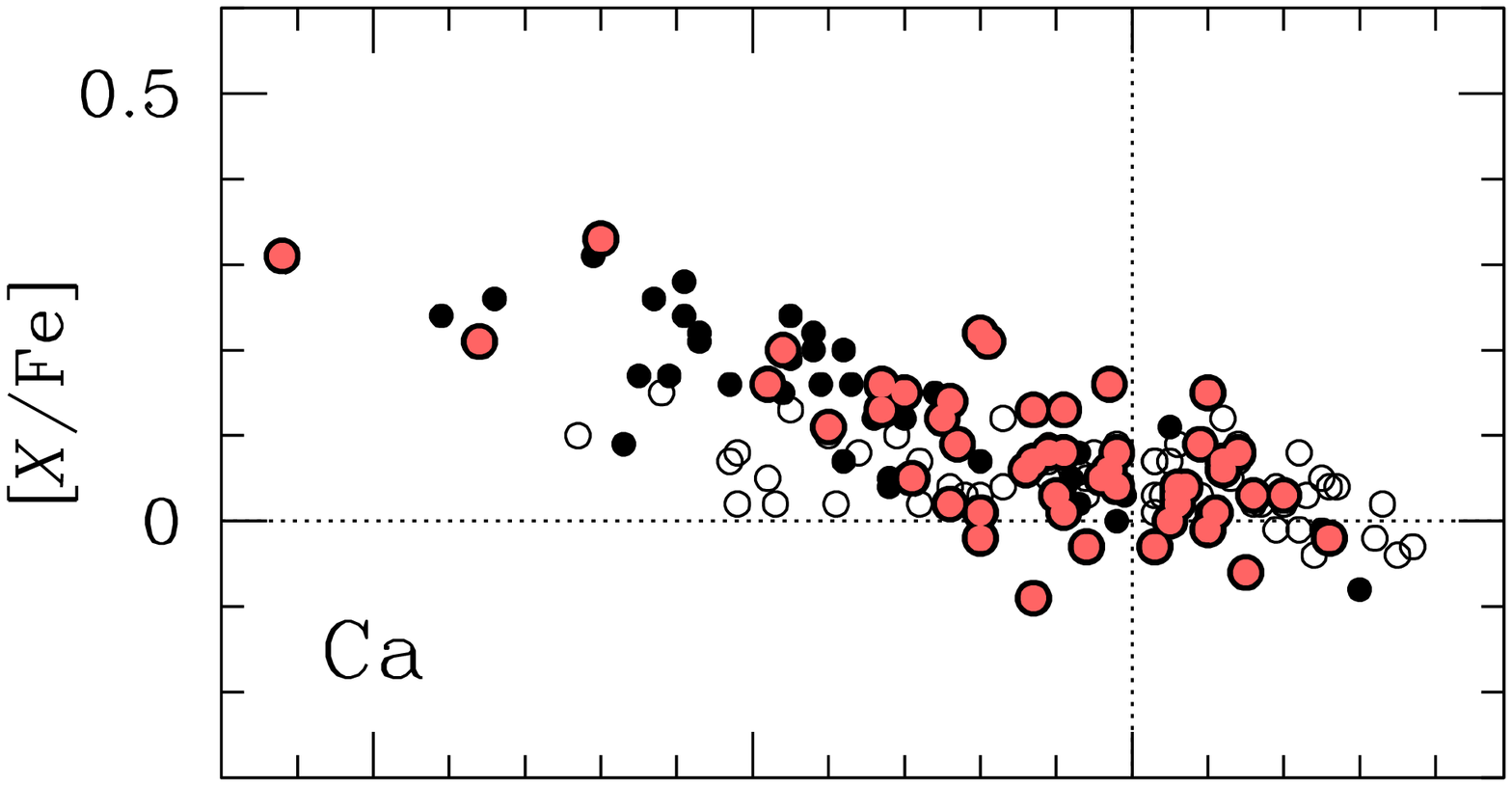}}
\resizebox{\hsize}{!}{\includegraphics[bb=18 223 562 508,clip]{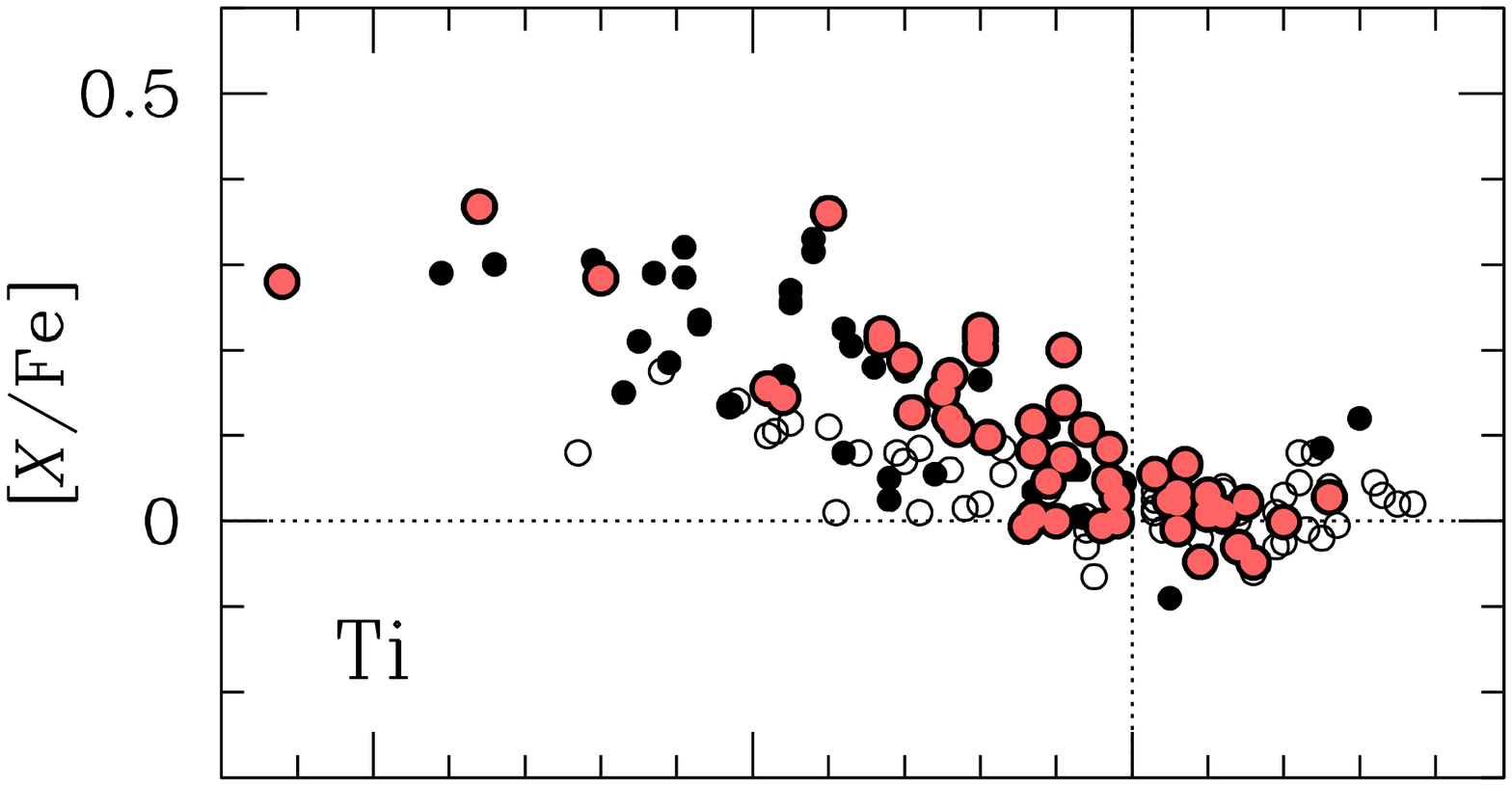}
                      \includegraphics[bb=90 223 562 508,clip]{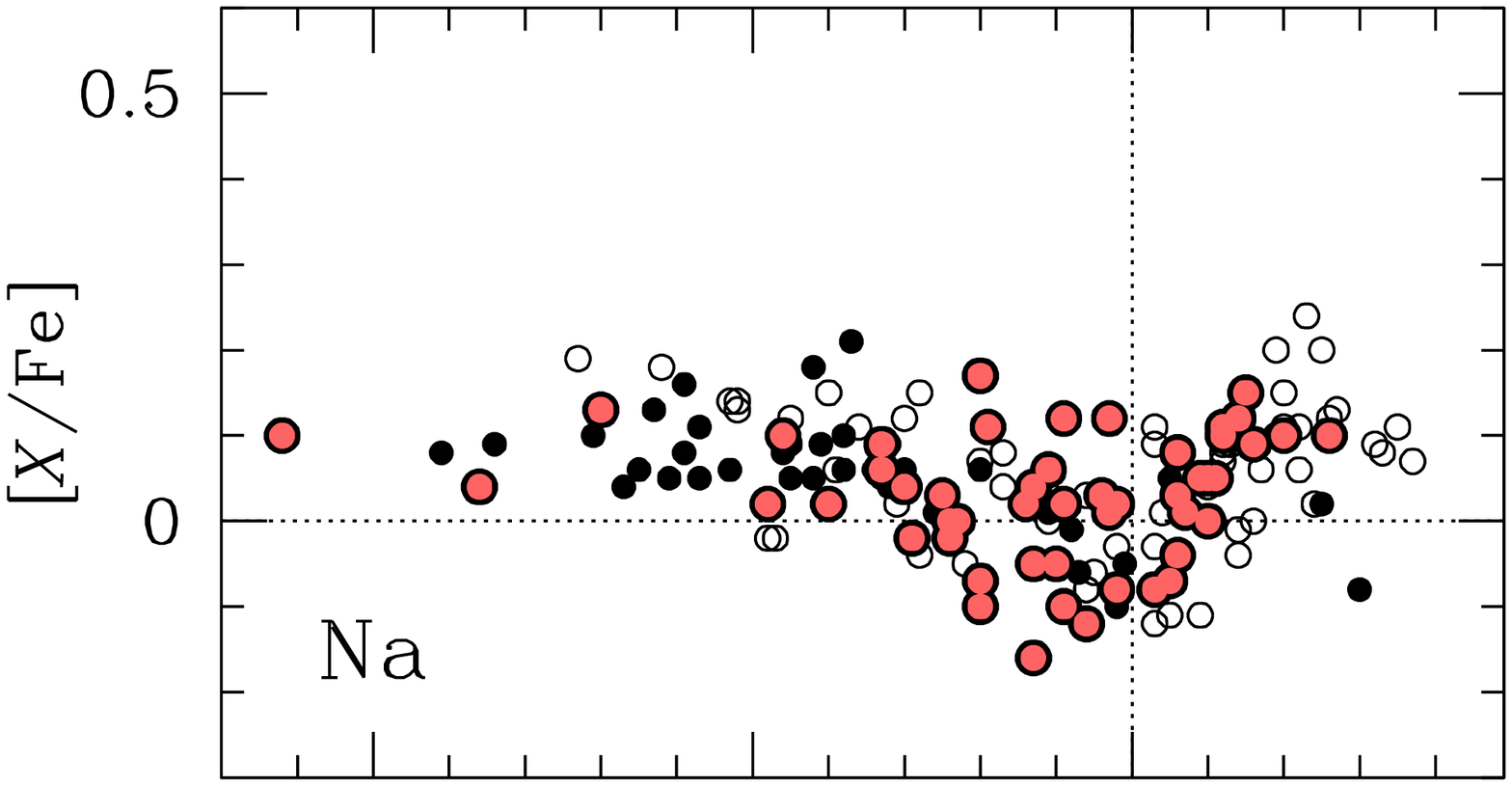}
                      \includegraphics[bb=90 223 592 508,clip]{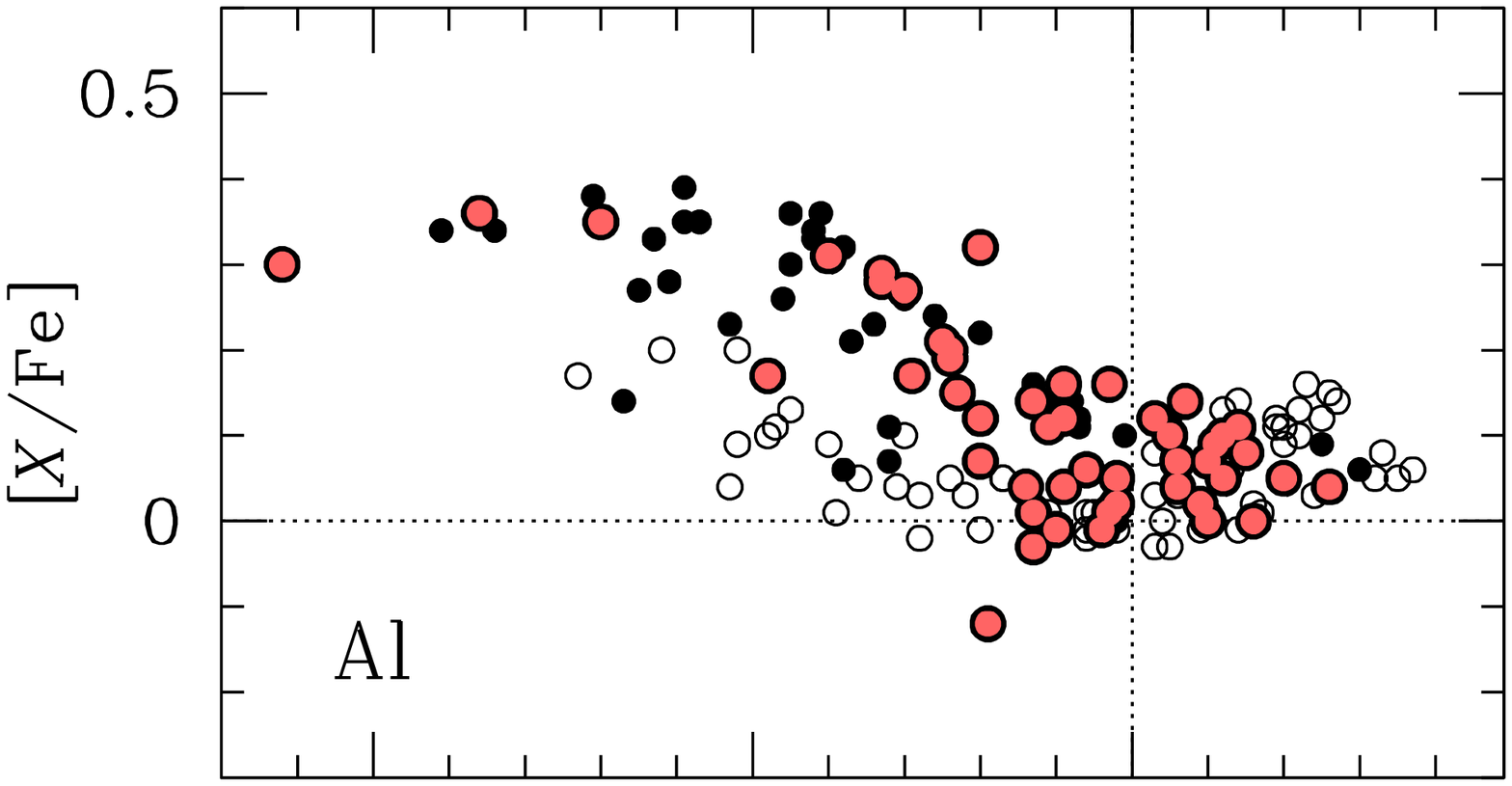}}
\resizebox{\hsize}{!}{\includegraphics[bb=18 154 562 508,clip]{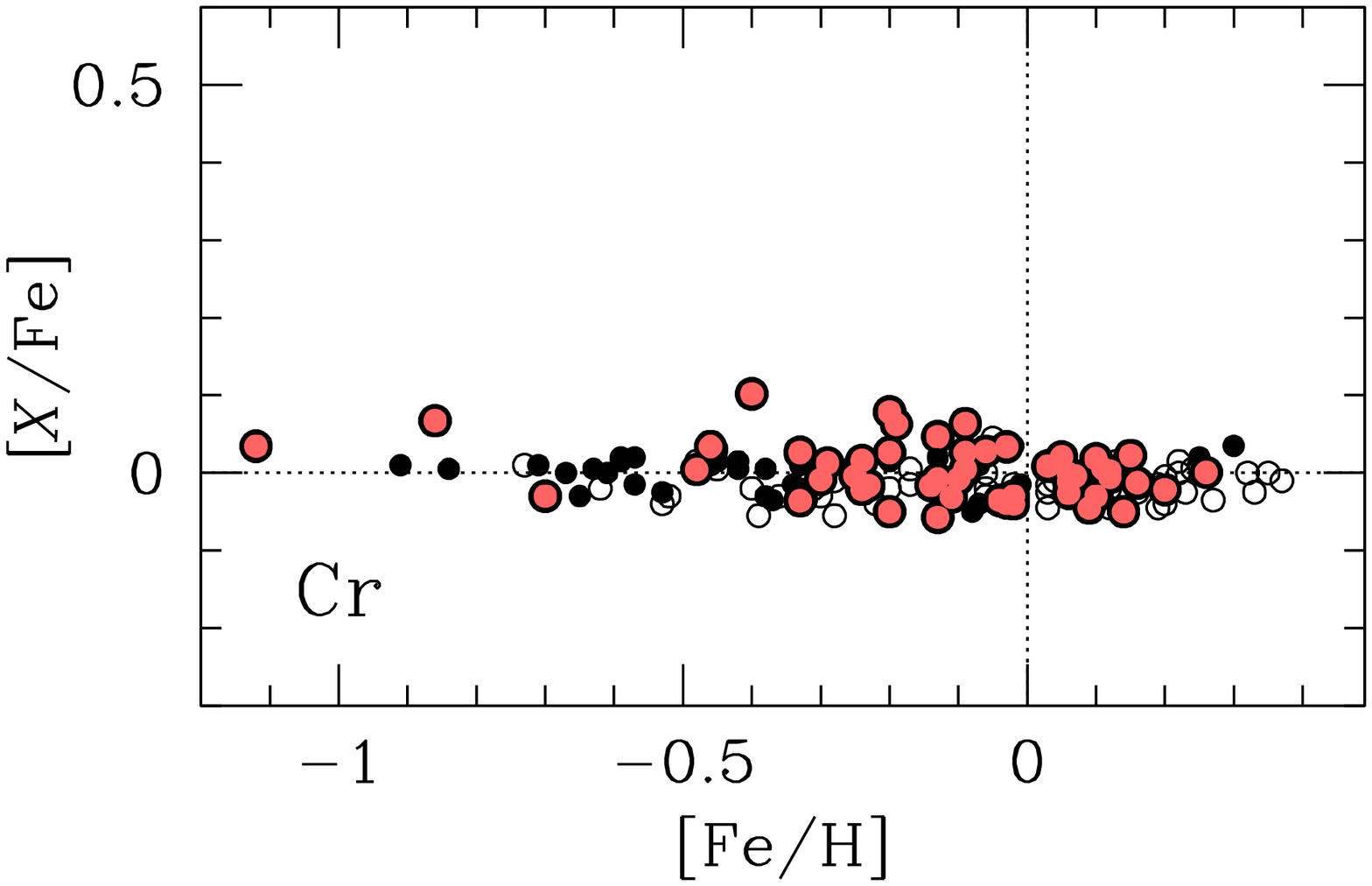}
                      \includegraphics[bb=90 154 562 508,clip]{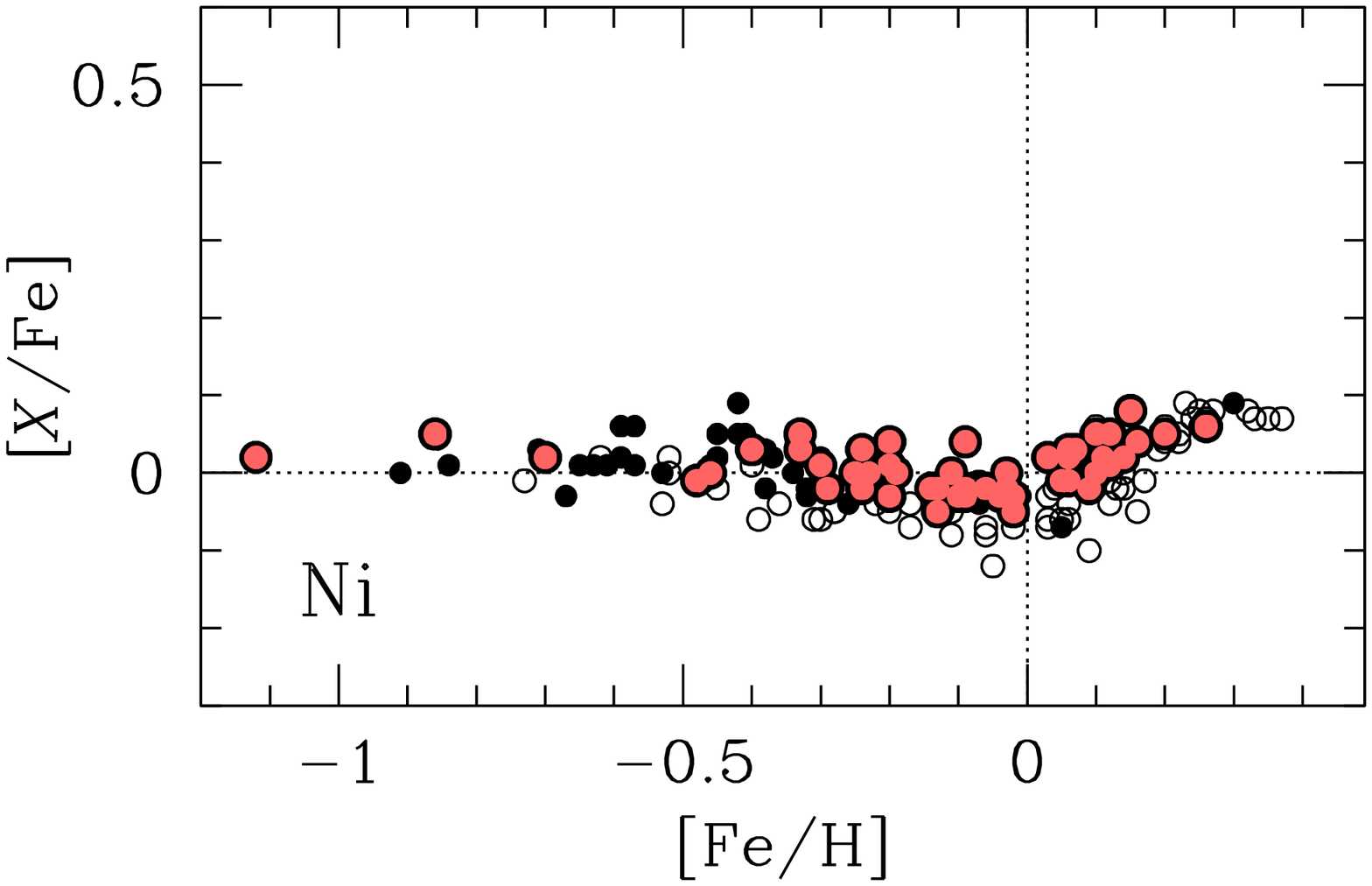}
                      \includegraphics[bb=90 154 592 508,clip]{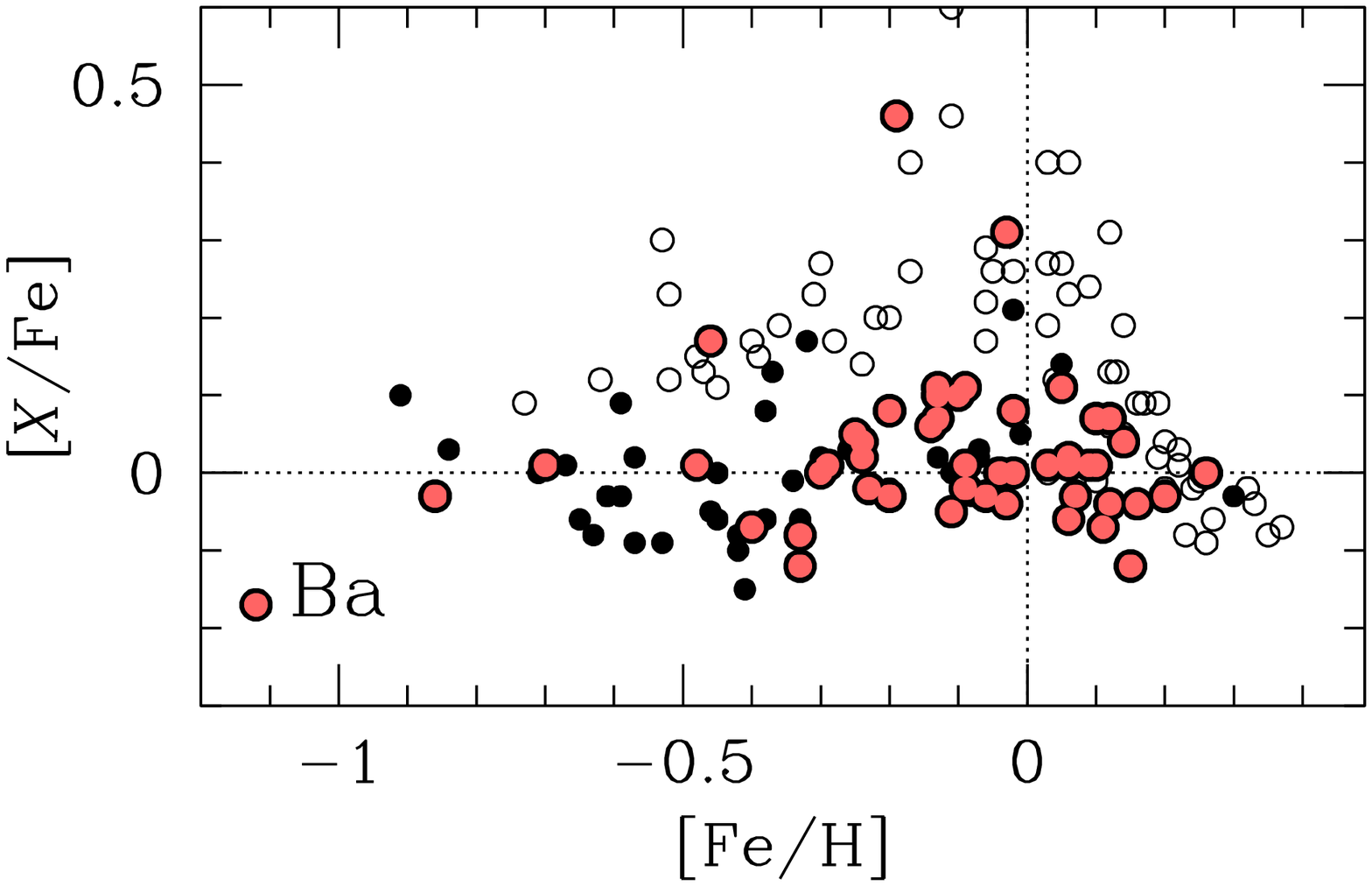}}

\caption{
        [{\it X}/Fe] vs. [Fe/H] abundance trends, where {\it X} is the
        element indicated in each plot.
        Hercules stream stars are marked by red (gray in B/W) 
        circles (gray in B/W). The thin and thick disk samples from
        \cite{bensby2003,bensby2005} are marked by open and black
        circles, respectively.
        }
\label{fig:abundancetrends}
\end{figure*}

\section{Results and discussion} \label{sec:results}

In Fig.~\ref{fig:toomre}b we plot ages versus [Fe/H] 
together with the thin and thick disk samples from 
\cite{bensby2003,bensby2005}. 
Below $\rm [Fe/H]=0$ it appears that the Hercules stream divides into two
(or maybe three) branches: one that follows the thick disk age
trend (the downward age-metallicity relation for the thick disk
was seen in \citealt{bensby_amr} and then also verified
by \citealt{haywood2006} and \citealt{schuster2006});
one that follows the thin disk age trend; and a
few stars (4-5) that tend to have high ages of $\sim15$\,Gyr
in the interval $\rm -0.4\lesssim[Fe/H]\lesssim0$. Although tempting, it
seems premature to conclude that we see a thin disk and a thick disk branch
in the Hercules stream. 

Abundance trends for nine different elements are shown in
Fig.~\ref{fig:abundancetrends}. Looking at the 
$\alpha$-elements (Mg, Si, Ca, Ti), the first impression is that 
most stars follow the trends as outlined by the thick disk stars.
At $\rm [Fe/H]\gtrsim-0.1$ the thin and thick disk are, however,
too close to distinguish in the $\alpha$-trends. But, the separation can be
extended for a few more tenths of a dex by looking at the Ba trend. Even here
most Hercules stars follow the thick disk trend. We note that Al also is a
good separator between the two disks; here we find that the trend for the
Hercules stars again accompanies the thick disk stars
The iron-peak element trends
are all remarkably tight, but add no more than a consistency check for
the Hercules stream abundance trends. 

Even though the Hercules stars do not split up as nicely in the
abundance plots as they did in the age-metallicity plot, they do indeed
mainly reproduce the trends as outlined by the thin and thick disks,
in fact, predominantly the thick disk. So what does this suggest about the
origin of stars of the Hercules stream?

{\bfseries Unique population?} The Hercules stream stars are
certainly unique in the sense that they form a well defined enhancement
in the local velocity distribution. Its age and abundance spreads, however,
tend to support that it is a mixture of populations.

{\bfseries Thin and thick disk mixture?}
For $\rm [Fe/H]\gtrsim -0.2$ there are several stars with thin disk
chemical compositions. The number of stars that follow the trends as
outlined by the thick disk are, however, more numerous. At super-solar
metallicities ($\rm [Fe/H]>0$) there is really no way to tell if we are 
looking at the thin or the thick disk. 

If the Hercules stream has a dynamical origin, 
consisting of stars originating at smaller galactocentric radii, are
we tracing the inner thin disk or the inner thick disk? As
no detailed abundance data for the inner disk(s) currently are available 
one can only speculate. Either it could be that there
is a substantial radial metallicity gradient in the thick disk
(and that is why we see so many metal-rich thick disk stars in the
Hercules stream), 
or that the chemical properties of the inner thin disk differs 
considerably from what we observe in the solar neighbourhood.
It is clear, however, that the Hercules stream divides into the
distinct abundance trends of the nearby thin and thick disks.

{\bfseries Bulge stars?}
This is not likely as recent studies find bulge stars to have 
large $\alpha$-enhancements at 
solar and super-solar [Fe/H] \citep{fulbright2006astroph,zoccali2006}. 

{\bfseries Recent merger?} 
It cannot be excluded that the stars of the
Hercules stream could originate from
a recent merger event between the Milky Way and another system. This system
must then have had properties very similar to the present properties
of the  Galactic thin and thick disks. Thus, such a hypothetical
merging galaxy would have chemical characteristics that depart considerably
from those of local dwarf galaxies \citep[cf, e.g.,][]{venn2004} and would 
presumably be more similar to a major spiral galaxy.

{\bfseries Conclusion:}
We conclude that the stars in the  Hercules stream seem to be a
mixture of thin and thick disk stars (especially the thick disk), 
supporting models that suggest that their kinematics are due to 
dynamical interactions with the Galactic bar. 
We note that these models suggest an inner disk origin for the Hercules
stream, and so we will need to compare our abundances with inner disk data, 
which we plan to obtain in the near future.

\acknowledgments

This work is supported by the National Science
Foundation, grant AST-0448900. SF is a Royal Swedish Academy
of Sciences Research Fellow supported by a grant from the Knut and
Alice Wallenberg Foundation.

\bibliographystyle{cupconf} 
\bibliography{referenser}

\begin{thebibliography}{26}
\expandafter\ifx\csname natexlab\endcsname\relax\def\natexlab#1{#1}\fi

\bibitem[{Arifyanto} \& {Fuchs}(2006)]{arifyanto2006}
{\sc {Arifyanto}, M.~I. \& {Fuchs}, B.} 2006, 
  {\em \aap\/} {\bf 449}, 533--538.

\bibitem[{Asplund} {\em et~al.\/}(2005){Asplund}, {Grevesse} \&
  {Sauval}]{asplundgrevessesauval2005}
{\sc {Asplund}, M., {Grevesse}, N. \& {Sauval}, A.~J.} 2005,
  In {\em ASP Conf. Ser. 336\/}, p.~25.

\bibitem[{Asplund} {\em et~al.\/}(1997){Asplund}, {Gustafsson}, {Kiselman} \&
  {Eriksson}]{asplund1997}
{\sc {Asplund}, M., {Gustafsson}, B., {Kiselman}, D. \& {Eriksson}, K.} 1997,
  {\em \aap\/} {\bf 318}, 521--534.

\bibitem[{Bensby} {\em et~al.\/}(2003){Bensby}, {Feltzing} \& {Lundstr{\"
  o}m}]{bensby2003}
{\sc {Bensby}, T., {Feltzing}, S. \& {Lundstr{\" o}m}, I.} 2003,
  {\em \aap\/} {\bf 410}, 527--551.

\bibitem[{Bensby} {\em et~al.\/}(2004){Bensby}, {Feltzing} \& {Lundstr{\"
  o}m}]{bensby_amr}
{\sc {Bensby}, T., {Feltzing}, S. \& {Lundstr{\" o}m}, I.} 2004,
  {\em \aap\/} {\bf 421},
  969--976.

\bibitem[{Bensby} {\em et~al.\/}(2005){Bensby}, {Feltzing}, {Lundstr{\" o}m} \&
  {Ilyin}]{bensby2005}
{\sc {Bensby}, T., {Feltzing}, S., {Lundstr{\" o}m}, I. \& {Ilyin}, I.} 2005,
  {\em \aap\/} {\bf 433}, 185--203.

\bibitem[{Bernstein} {\em et~al.\/}(2003){Bernstein}, {Shectman}, {Gunnels},
  {Mochnacki} \& {Athey}]{bernstein2003}
{\sc {Bernstein}, R., {Shectman}, S.~A., {Gunnels}, S.~M., {Mochnacki}, S. \&
  {Athey}, A.~E.} 2003,
  In {\em Proceedings of the SPIE,
  Volume 4841\/} (ed. M.~{Iye} \& A.~F.~M. {Moorwood}), pp. 1694--1704.

\bibitem[{Dehnen}(1998)]{dehnen1998a}
{\sc {Dehnen}, W.} 1998,
  {\em \aj\/} {\bf 115}, 2384--2396.

\bibitem[{Dehnen}(1999)]{dehnen1999}
{\sc {Dehnen}, W.} 1999,
 {\em \apjl\/}
  {\bf 524}, L35--L38.

\bibitem[{Dehnen}(2000)]{dehnen2000}
{\sc {Dehnen}, W.} 2000,
  {\em \aj\/} {\bf
  119}, 800--812.

\bibitem[{Demarque} {\em et~al.\/}(2004){Demarque}, {Woo}, {Kim} \&
  {Yi}]{demarque2004}
{\sc {Demarque}, P., {Woo}, J.-H., {Kim}, Y.-C. \& {Yi}, S.~K.} 2004,
  {\em \apjs\/} {\bf
  155}, 667--674.

\bibitem[{Ecuvillon} {\em et~al.\/}(2006){Ecuvillon}, {Israelian}, {Pont},
  {Santos} \& {Mayor}]{ecuvillon2006astroph}
{\sc {Ecuvillon}, A., {Israelian}, G., {Pont}, F., {Santos}, N.~C. \& {Mayor},
  M.} 2006,
  {\em \aap, accepted,
  (astro-ph/0608669)\/} .

\bibitem[{Edvardsson} {\em et~al.\/}(1993){Edvardsson}, {Andersen},
  {Gustafsson}, {Lambert}, {Nissen} \& {Tomkin}]{edvardsson1993}
{\sc {Edvardsson}, B., {Andersen}, J., {Gustafsson}, B., {Lambert}, D.~L.,
  {Nissen}, P.~E. \& {Tomkin}, J.} 1993,
  {\em \aap\/} {\bf 275}, 101--+.

\bibitem[{ESA}(1997)]{esa1997}
{\sc {ESA}} 1997 {\em {The HIPPARCOS and TYCHO catalogues, ESA SP Ser. vol. 1200, Noordwijk}\/}.

\bibitem[{Famaey} {\em et~al.\/}(2005){Famaey}, {Jorissen}, {Luri}, {Mayor},
  {Udry}, {Dejonghe} \& {Turon}]{famaey2005}
{\sc {Famaey}, B., {Jorissen}, A., {Luri}, X., {Mayor}, M., {Udry}, S.,
  {Dejonghe}, H. \& {Turon}, C.} 2005,
  {\em \aap\/} {\bf 430}, 165--186.

\bibitem[{Fulbright} {\em et~al.\/}(2006){Fulbright}, {McWilliam} \&
  {Rich}]{fulbright2006astroph}
{\sc {Fulbright}, J.~P., {McWilliam}, A. \& {Rich}, R.~M.} 2006,
  {\em \aj, submitted, (astro-ph/0609087)\/} .

\bibitem[{Fux}(2001)]{fux2001}
{\sc {Fux}, R.} 2001,
  {\em \aap\/} {\bf 373}, 511--535.

\bibitem[{Gustafsson} {\em et~al.\/}(1975){Gustafsson}, {Bell}, {Eriksson} \&
  {Nordlund}]{gustafsson1975}
{\sc {Gustafsson}, B., {Bell}, R.~A., {Eriksson}, K. \& {Nordlund}, A.} 1975,
  {\em \aap\/}
  {\bf 42}, 407--432.

\bibitem[{Haywood}(2006)]{haywood2006}
{\sc {Haywood}, M.} 2006,
  {\em \mnras\/} {\bf 371}, 1760--1776.

\bibitem[{Helmi} {\em et~al.\/}(2006){Helmi}, {Navarro}, {Nordstr{\"o}m},
  {Holmberg}, {Abadi} \& {Steinmetz}]{helmi2006}
{\sc {Helmi}, A., {Navarro}, J.~F., {Nordstr{\"o}m}, B., {Holmberg}, J.,
  {Abadi}, M.~G. \& {Steinmetz}, M.} 2006,
  {\em \mnras\/} {\bf 365}, 1309--1323.

\bibitem[{Kim} {\em et~al.\/}(2002){Kim}, {Demarque}, {Yi} \&
  {Alexander}]{kim2002}
{\sc {Kim}, Y., {Demarque}, P., {Yi}, S.~K. \& {Alexander}, D.~R.} 2002,
  {\em \apjs\/}
  {\bf 143}, 499--511.

\bibitem[{Raboud} {\em et~al.\/}(1998){Raboud}, {Grenon}, {Martinet}, {Fux} \&
  {Udry}]{raboud1998}
{\sc {Raboud}, D., {Grenon}, M., {Martinet}, L., {Fux}, R. \& {Udry}, S.} 1998,
  {\em \aap\/} {\bf 335}, L61--L64.

\bibitem[{Schuster} {\em et~al.\/}(2006){Schuster}, {Moitinho}, {M{\'a}rquez},
  {Parrao} \& {Covarrubias}]{schuster2006}
{\sc {Schuster}, W.~J., {Moitinho}, A., {M{\'a}rquez}, A., {Parrao}, L. \&
  {Covarrubias}, E.} 2006,
  {\em \aap\/} {\bf
  445}, 939--958.

\bibitem[{Skuljan} {\em et~al.\/}(1999){Skuljan}, {Hearnshaw} \&
  {Cottrell}]{skuljan1999}
{\sc {Skuljan}, J., {Hearnshaw}, J.~B. \& {Cottrell}, P.~L.} 1999,
  {\em \mnras\/} {\bf 308},
  731--740.

\bibitem[{Venn} {\em et~al.\/}(2004){Venn}, {Irwin}, {Shetrone}, {Tout}, {Hill}
  \& {Tolstoy}]{venn2004}
{\sc {Venn}, K.~A., {Irwin}, M., {Shetrone}, M.~D., {Tout}, C.~A., {Hill}, V.
  \& {Tolstoy}, E.} 2004,
  {\em \aj\/} {\bf 128}, 1177--1195.

\bibitem[{Zoccali} {\em et~al.\/}(2006){Zoccali}, {Lecureur}, {Barbuy}, {Hill},
  {Renzini}, {Minniti}, {Momany}, {Gomez} \& {Ortolani}]{zoccali2006}
{\sc {Zoccali}, M., {Lecureur}, A., {Barbuy}, B., {Hill}, V., {Renzini}, A.,
  {Minniti}, D., {Momany}, Y., {Gomez}, A. \& {Ortolani}, S.} 2006,
  {\em \aap\/} {\bf 457}, L1--L4.

\end{thebibliography}

\end{document}